# Polarization-dependent light extinction in ensembles of polydisperse, vertical semiconductor nanowires: A Mie scattering effective medium

**Grzegorz Grzela (1), Djamila Hourlier (2), Jaime Gómez Rivas (1,3)**

(1) FOM-Institute AMOLF, c/o Philips Research Laboratories Eindhoven, High Tech Campus 4, 5656 AE Eindhoven, The Netherlands

(2) Institut d'Electronique, de Microelectronique et de Nanotechnologies, UMR-CNRS 8520, F-59652 Villeneuve d'Ascq, France

(3) COBRA Research Institute, Eindhoven University of Technology, P.O. Box 513, 5600 MB Eindhoven, The Netherlands

## Abstract

We present an experimental and theoretical study of the angle- and polarization-dependent light extinction in random arrays of polydisperse semiconductor nanowires epitaxially grown on substrates. The specular reflectance is described by averaging the scattering properties of individual nanowires obtained from Lorenz-Mie theory over the diameter distribution. The complex effective refractive index describing the propagation and attenuation of the coherent beam scattered in forward direction is determined in the independent scattering approximation and used to calculate the angle- and polarization-dependent reflectance. Our measurements demonstrate the highly anisotropic scattering in ensembles of aligned nanowires.

# 1. Introduction

Progress in the growth and doping of semiconductor nanowires has led to an increased interest on these nanostructures. This interest has been driven by the promising perspectives of nanowires in applications such as nano-light-emitting diodes,[1] lasers,[2,3] single-photon sources[4] and solar cells.[5] On a more fundamental level, semiconductor nanowires are also being intensively investigated for their unique optical properties. The optical anisotropy in the polarized emission and absorption by individual nanowires, associated to their large geometrical anisotropy, was first reported and described by Wang et al.[6] However, this description was based on the quasistatic approximation, which assumes a uniform electromagnetic field amplitude over the cross section of the nanowire. At optical frequencies, this approximation is valid for nanowires with a diameter smaller than 10 nm.[7] Optical retardation in nanowires with larger diameter requires a more elaborated modeling of their optical properties. Van Weert et al. suggested that the Lorenz-Mie scattering theory for infinitely long circular cylinders could be applied to describe more precisely the absorption and emission anisotropies of semiconductor nanowires.[8] Mie scattering theory has been used by Cao et al. to describe the resonant enhancement of the absorption by individual nanowires.[9] This enhancement has been also proposed to increase the efficiency of nanowire-based thin film solar cells.[10] Most of the suggested applications will rely on periodic and random arrays of nanowires instead of on single nanostructures. Therefore, the description of the light propagation, scattering and absorption in these arrays is becoming increasingly important. Recent achievements in this direction are the demonstration of giant birefringence in dense arrays of nanowires,[11] the demonstration of graded index layers with minimum reflection formed by nanowires grown on top of substrates,[12, 13] the proposal of photonic crystal nanowire structures for enhanced absorption in photovoltaic applications,[14-17], and the description of multiple scattering in random mats of semiconductor nanowires.[18-20]

In this article, we use the Mie-theory in the independent scattering approximation to describe the anisotropic scattering and extinction of light in a random ensemble of epitaxially grown Si nanowires. This analytical model allows us to determine the effective index of refraction and extinction coefficients, describing the propagation of the beam scattered in the forward direction. This beam is also called the coherent beam in contrast to the diffuse light that results from multiple scattering in the sample.[21] The approach is valid in the low scatter-density limit, which is the case for the investigated sample that consists of nanowires with a volume filling fraction of $\alpha \cong 0.025$. The real component of the effective refractive index is close to unity as expected from the low nanowire filling fraction. The imaginary component, defining the extinction length, is highly anisotropic and wavelength dependent. At long wavelengths and large angles of incidence the extinction length for s-polarization is more than three times larger than for p-polarization and solely determined by scattering in the nanowire layer. The calculated specular reflectance using the effective medium optical constants is in excellent agreement with the measurements. Our demonstration of the validity of Mie effective medium theory to describe the properties of complex arrays of polydisperse nanowires facilitates the design of light scattering and absorption in ensembles of nanowires for, e.g., photovoltaic applications.

## 2. Effective refractive index

Dielectric mixing rules provide the value of the effective dielectric constants, e.g., the complex refractive index $\tilde{n}_{eff} = n_{eff} + ik_{eff}$, characterizing the electromagnetic response of heterogeneous materials formed by two or more different constituents. These effective medium constants describe the phase and amplitude of an optical beam propagating through the medium as if it was homogeneous with a response that it is given by that of its constituents and their filling fraction. If the medium is formed by particles or scatterers with a size much smaller than the wavelength λ of light, they create a perturbation of the field that in the lowest order is that of a dipole. In this case, the effective refractive index is calculated in the quasistatic approximation. The simplest mixing formula in the quasistatic limit is the well-known Maxwell-Garnett formula, which considers small dipolar spheres embedded in a matrix.

As the size of the particles increases, their optical response cannot be approximated by that of dipolar spheres. To calculate the effective medium response, the multipolar response of individual particles in the ensemble must be considered. The far field radiation scattered by a particle of dimension $r$ in any direction defined by the angles Θ and Φ in spherical coordinates, when illuminated by light of wavelength λ from a direction defined by the angles θ and φ, is described by the amplitude scattering function $S(\Theta, \Phi, r, \theta, \varphi, \lambda)$ of the particle. This complex scattering function contains information about the amplitude (real component) and phase (imaginary component) of the scattered wave. The extinction of light, which is the result of the scattering and absorption, is defined as the decrease of the electric field amplitude in the direction of propagation. Therefore, the optical extinction produced by a scatterer is related to the amplitude scattering function at $\Theta = \pi + \theta$ and $\Phi = \pi + \varphi$, which for simplicity we will denote as $S_0(r, \theta, \varphi, \lambda)$.

The optical extinction in an ensemble of finite particles in the independent scattering approximation is given by the sum of the amplitude scattering functions of the individual particles. The independent scattering approximation is valid for scattering samples with a low density of scatterers. This approximation neglects the probability that the scattered light by one particle returns to the particle after being scattered somewhere else in the sample. As we will show later, this condition is fulfilled by our sample with a nanowire volume filling fraction of 0.025. For an ensemble of polydisperse scatterers, it is necessary to weight the amplitude scattering functions using the normalized particle size distribution function $f(r)$ describing the polydispersity

$$\overline{S_0(\theta, \varphi, \lambda)} = \int_0^\infty f(r) S_0(r, \theta, \varphi, \lambda) dr. \quad (1)$$

The propagation of the coherent beam is given by the sum of the scattered amplitudes in the forward direction from the individual particles. This propagation can be described as the propagation in a homogeneous medium with an effective complex refractive index given by,[22]

$$\tilde{n}_{eff}(\theta, \lambda) = 1 + i \frac{2\pi \rho_V}{k^3} \overline{S_0(\theta, \varphi, \lambda)}, \quad (2)$$

where $\rho_V$ is the density of scatterers and $k = 2\pi/\lambda$ is the wave number in vacuum. Therefore, the determination of the effective refractive index for the coherent beam is reduced to the determination of the amplitude scattering function of the scatterers. The real component of this index defines the phase delay or advance in the propagation of the coherent beam through the scattering medium. The imaginary component defines the effective extinction coefficient that describes the attenuation of the amplitude in the forward direction due to absorption and scattering. The spatial dispersion in the heterogeneous medium, introduced by the scatterers, gives rise to an anisotropic response that translates into an angle-dependent effective index.

The scattering properties of nanowires can be calculated using the Lorenz-Mie theory.[23] This theory describes the light scattering by an individual, infinitely long circular cylinder. By expanding the incident and scattered fields into cylindrical harmonics, it is possible to calculate the amplitude scattering function for such a scatterer. The amplitude scattering function depends on the radius of the cylinder $r$, the angles of incidence $\theta$ and $\varphi$, the wavelength $\lambda$ and the polarization of the incident light, which can be defined parallel ($\parallel$) or perpendicular ($\perp$) to the axis of the cylinder. To distinguish the amplitude scattering function in the forward direction of an infinitely long cylinder from its counterpart of a particle, we denote it as $T_{0,\parallel,\perp}(r, \theta, \varphi, \lambda)$.

Semiconductor nanowires have a finite length. For nanowire lengths much larger than their diameter, $l >> 2r$, and at long distances from the nanowire, $z >> l^2/\lambda$, the amplitude scattering function of nanowires with a finite length be approximated to the Mie amplitude scattering function of infinitely long cylinders. In that case, the amplitude scattering function in the forward direction of finite particles is related to the amplitude scattering function in forward direction of infinite cylinders by[22]

$$\overline{S_{0,\parallel,\perp}(\theta,\varphi,\lambda)} \cong \frac{kl}{\pi}\overline{T_{0,\parallel,\perp}(\theta,\varphi,\lambda)}, \qquad (3)$$

where $\overline{T_{0,\parallel,\perp}(\theta,\varphi,\lambda)}$ is the weighted Mie scattering function in the forward direction for cylinders. The nonweighted amplitude scattering functions for both polarizations ($\parallel, \perp$) are infinite series of the scattering coefficients:

$$T_{0,\parallel}(r,\theta,\varphi,\lambda) = b_{0,\parallel} + 2\sum_{n=1}^{\infty} b_{n,\parallel},$$

$$T_{0,\perp}(r,\theta,\varphi,\lambda) = a_{0,\perp} + 2\sum_{n=1}^{\infty} a_{n,\perp}.$$

(4)

The scattering coefficients are functions of the cylinder's radius, angle of incidence, wavelength and material constants. Explicit mathematical expressions of these coefficients can be found in Ref. [23].

The effective complex refractive index for the p- and s-polarized coherent beam propagating in an ensemble of vertically aligned nanowires is given by

$$\tilde{n}_{eff,p,s}(\theta,\lambda) \cong 1 + i\frac{2\rho_V l}{k^2}\overline{T_{0,\parallel,\perp}(\theta,\varphi,\lambda)}, \qquad (5)$$

where the density of scatterers is related to the volume filling fraction $\alpha$ by

$$\rho_V = \frac{N}{V} = \frac{\alpha}{\pi l} \frac{1}{\int_0^\infty f(r) r^2 dr}. \tag{6}$$

Using this approach, different scatterers can be included in the calculations, such as cylinders with a random and nonvertical orientation. In this way, the additional scatterers have an influence on the real and imaginary components of the effective refractive index. If the polarization-, orientation- and size-averaged amplitude scattering function of nonvertical nanowires is denoted as $\overline{T_0(\lambda)}$, the effective refractive index is given by

$$\tilde{n}_{eff,p,s}(\theta, \lambda) \cong \\ 1 + i \frac{\lambda^2}{2\pi^3} \frac{1}{\int_0^\infty f(r) r^2 dr} \left( \alpha_{vertical} \overline{T_{0,\parallel,\perp}(\theta, \varphi, \lambda)} + \alpha_{nonvertical} \overline{T_0(\lambda)} \right), \tag{7}$$

where $\alpha_{nonvertical}$ is the volume filling fraction of nonvertical cylinders.

The specular reflectance of the system consisting of the nanowire layer and the semi-infinite Si substrate can be calculated using the transfer matrix method.[24] This method calculates the reflection and transmission coefficients of a multilayered structure. A polarization independent scattering term can be introduced to account for a reduction of the specular reflection on the interfaces due to surface scattering. The specular reflectance at an interface in the presence of surface roughness scattering $R_s$ is given by[25]

$$R_s = R_0 e^{-\left(\frac{4\pi\sigma \cos\theta}{\lambda}\right)^2}, \tag{8}$$

where $R_0$ is the specular reflectance in the absence of surface scattering and $\sigma$ is the root mean square height of the surface features causing the roughness.

## 3. Sample description

We have grown silicon nanowires on top of a silicon substrate using the vapor-liquid-solid (VLS) method in a chemical vapor deposition reactor.[26] A gold layer with a nominal thickness of 5 nm was evaporated on a single-side polished, n-type silicon (111) substrate. The gold-coated substrate was heated to temperature T=650°C under vacuum to catalyze the growth of Si nanowires. The nanowires were grown by pyrolyzing silane diluted in 10% $H_2$ in argon under a gas flow dynamics and a pressure ratio $P_{SiH4}/P_{H2}$ = 33x10$^{-3}$. The growth resulted in a random polydisperse ensemble of silicon nanowires. Figure 1 (a) shows a cross-section scanning-electron microscope (SEM) image of a layer of Si nanowires with an average length of 7.3 µm. A SEM image of the tilted sample is shown in Fig. 1 (b). On top of each nanowire the catalyst particle is clearly visible. The nanowires are aligned preferentially in the (111) direction, with 80% of the wires in the sample in this vertical direction. The overall volume filling fraction of silicon in form of vertical nanowires was estimated to be $\alpha_{vertical} \cong 0.02$. There is also a small fraction of non-vertical nanowires with a volume fraction $\alpha_{nonvertical} \cong 0.005$. Also in Fig. 1 (a) and (b) a significant surface roughness can be observed on the interface between the nanowire layer and the underlying substrate. As we will show later, this roughness leads to additional light scattering and a reduction of the specular reflection.

The diameter distribution of the nanowires was determined from several SEM images taken on the sample at different places. The histogram displaying this distribution is shown in Fig. 1 (c). The diameter distribution can be approximated with a log-normal probability density function with a mean of 182 nm and standard deviation 0.4 nm. This probability density function is represented by the black solid line in Fig 1 (c). The area under the probability density function was normalized to unity for the calculations. The upper inset of Fig. 1 (c) is a photograph of the sample where the brownish area is occupied by nanowires and the small dark area on the right side of the sample is the silicon wafer where no nanowires were grown. The region with the nanowires has a matte appearance characteristic of scattering samples.

## 4. Results and discussion

To explore the propagation of the coherent beam through the ensemble of nanowires, the specular reflectance as a function of the wavelength and the angle of incidence was measured for p- and s-polarized light using a coherent supercontinuum light source. This light source provided enough optical power to distinguish with the eye between specularly and diffusively reflected light. A schematic representation of the specular reflectance measurement configuration is displayed as the lower inset of Fig. 1(c). The angle of incidence is defined with respect to the normal of the sample and coincides with the angle $\theta$ defining the amplitude scattering function of vertically aligned nanowires. Figures 2 (a) and (b) show the measurement for s- and p-polarization respectively. Reflection measurements, instead of transmission, were done to extend the measurement range to higher energies than the band gap of silicon at which absorption in the substrate suppresses the transmission. Fabry-Perót oscillations are visible due to the finite thickness of the nanowire layer and the interference of the light reflected at its upper and lower interfaces. The magnitude of the specular reflectance is very low, below 0.015, which indicates a very large optical extinction in the sample. The reflectance for both polarizations increases for small angles due to the reduction of the optical path length in the scattering layer. At wavelengths longer than the Si band gap wavelength ($\lambda$=1100 nm) the extinction is caused by scattering out of the beam propagating in the forward direction, while at shorter wavelengths the nanowires can scatter and absorb the intensity of this beam.

To exclude a possible anisotropic response of the sample with respect to the azimuthal angle $\varphi$, we measured the specular reflectance at $\lambda$=633 nm at normal incidence as a function of this angle (Fig. 3). Because of the cylindrical symmetry of vertical nanowires along their axis, they have an isotropic response when illuminated at normal incidence. Therefore, any possible anisotropic response at this angle of incidence could be attributed to the small fraction of nonvertical nanowires. As can be appreciated in Fig. 3, the reflectance does not vary with $\varphi$, which indicates that the nonvertical nanowires are not preferentially aligned along any azimuthal angle. Furthermore, we assume that the nonvertical nanowires have also a random inclination in order to average the amplitude scattering function over $\theta$ and the polarization.

Since the thickness, the fraction of vertical and nonvertical nanowires and the diameter distribution of the nanowires have been determined from the SEM images, we can calculate the effective complex refractive index of the layer describing the propagation and attenuation of the coherent beam by using the Lorenz-Mie model. We consider here two cases: A layer consisting of polydisperse vertical nanowires with a volume filling fraction of $\alpha_{vertical} \cong 0.02$ and a layer in which there is an additional volume fraction $\alpha_{nonvertical} \cong 0.005$ of randomly oriented nonvertical nanowires. For determining the angle-, wavelength- and polarization-dependent effective index we use Eq. (7), in which the amplitude scattering functions were calculated using the Lorenz-Mie formalism and weighted with the log-normal diameter distribution function shown in Fig. 1 (c). As input parameters we use the radii of the cylinders, angle of incidence, wavelength. and the wavelength-dependent refractive index of crystalline silicon obtained from Ref [27].[27] Figure 4 shows the real and imaginary components of the calculated effective refractive indices for p-polarized (solid and dash-dotted black lines) and s-polarized (dashed and dotted red lines) light and for three different wavelengths, i.e., $\lambda$=633 nm (a,b), $\lambda$=850 nm (c,d) and $\lambda$=1200 nm (e,f). The solid and dotted lines represent the calculations for the layer of vertical nanowires, whereas the dashed and dash-dotted lines stand for the ensemble consisting of both vertical and nonvertical nanowires. The effective complex refractive index is the same for both polarizations at $\theta$ = 0°, where the polarization is indistinguishable because of the cylindrical symmetry of nanowires. The real part of the effective refractive index is close to unity for both polarizations due to the very low volume filling fraction of the nanowires. Such a low refractive index facilitates broadband and omnidirectional light coupling into the nanowire layer.[28] The resonant behavior of Mie scatterers at wavelengths comparable to their diameter leads to an effective refractive index smaller than unity at short wavelengths and mainly for p-polarized light. The refractive index that is close to, but smaller than, unity means a phase advance of the wave propagating in the forward direction through the low-density medium of resonant scatterers. An elegant description of this phenomenon can be found in chapter 9 of Ref. [23]. The imaginary component of the effective refractive index, or the effective extinction coefficient, decreases with the angle of incidence because of the reduction of geometrical cross section of the nanowires. Due to larger extinction efficiency of individual nanowires for p-polarized light, the effective extinction coefficients of the ensembles of nanowires are also larger for that polarization compared to s-polarization. This fact, together with the lower Fresnel reflection at the interfaces for p-polarized light, is responsible for the large differences in intensity between s- and p-polarized light reflected from the sample. The addition of nonvertical nanowires in the layer modifies the complex effective refractive index. In the case of the real component, the influence of the non-vertical nanowires is small due to the low density of the nonvertical nanowires. This influence is much more evident for the effective extinction coefficient.

The calculated effective refractive index can be used to compute the specular reflectance of the coherent beam in the nanowire layer by applying the transfer matrix method. Three layers are considered for these calculations, i.e., a semi-infinite layer of air (n = 1). from which the light is incident; an effective medium for the coherent beam describing the layer of nanowires with a thickness of 7.3 µm; and a semi-infinite substrate of silicon on top of which the nanowires are grown. In Fig. 5 we display the specular reflectance measurements at the three different wavelengths, i.e., $\lambda$=633, 850, and 1200 nm, and for s- (left column) and p-polarization (right column) together with the transfer matrix calculations. The measurements are represented by symbols (red circles for s-polarization and black circles for p-polarization), whereas the calculations are illustrated with the solid lines. If only the vertical nanowires are considered, the calculated specular reflectance (black dotted lines) is significantly higher than the measurement, though there is a qualitative agreement in the Fabry-Perót oscillations. The addition of the isotropic extinction caused by the randomly oriented nonvertical nanowires provides a better quantitative agreement between the calculated (red dash-dotted lines) and the measured reflectance. In addition, we can correct the calculations to take into account the extinction due to surface roughness scattering by using the Eq. (8). The root mean square feature height $\sigma$, used to fit the measurements for all wavelengths and polarizations is 60 nm. This value is reasonable to effectively take into account surface irregularities of the bottom and upper interface of the nanowire layer. The possible light extinction by the gold particles on top of each nanowire is included into the surface roughness scattering. Therefore, the value of $\sigma$ might overestimate the surface roughness of the sample. As can be appreciated in Fig. 5, the correction to the specular reflection introduced by the nonvertical nanowires and the surface roughness results in excellent agreement between the calculation (blue solid lines) and the measurements.

To illustrate the extinction anisotropy of an ensemble of vertical nanowires, we derived the extinction lengths for s- and p-polarized light obtained from the measurements of the extinction coefficient by knowing that $l_{ext} = \lambda/4\pi k_{eff}$. Figure 6 shows the ratio of extinction lengths, $l_{ext,s}/l_{ext,p}$. We present this ratio at the three different wavelengths, i.e., $\lambda$=633 nm (solid line), $\lambda$=850 nm (dashed line) and $\lambda$=1200 nm (dotted line) as a function of the angle of incidence. As expected, the anisotropy vanishes for normally incident light. Because of the strong polarization dependence of the light scattering and absorption by nanowires,[29] the layer consisting of vertical nanowires shows large extinction anisotropy for larger angles of incident light. These measurements represent the first demonstration of anisotropic scattering and absorption in random ensembles of semiconductor nanowires. For the wavelengths in the absorbing range of silicon, such as λ=633 nm, the extinction length of s-polarized light can be nearly twice as long as that for p-polarized light in our sample. The ratio of extinction lengths in the near-infrared regime is even larger, achieving a value of 3.5 for λ=1200 nm and at large angles of incidence. For this wavelength the extinction is solely due to scattering.

## 5. Conclusions

Layers of nanowires form a medium with an effective complex refractive index that characterizes the propagation and extinction of the coherent beam or the beam propagating in the forward direction. We have derived the polarization-, angle of incidence- and wavelength-dependent refraction and extinction of the coherent beam in the ensembles of polydisperse silicon nanowires from the properties of individual infinite cylinders in the independent scattering approximation. Due to the low effective refractive index of the nanowire layer, the incident light can be easily transmitted into the nanowire layer, where it is extinct by scattering and absorption. Our measurements reveal a strongly anisotropic scattering and absorption in the nanowire layer. The description of light propagation and extinction in nanowire layers is of relevance for the design of novel nanowire solar cells in which scattering and absorption need to be optimized.

# Acknowledgements

This work is part of the research program of the "Stichting voor Fundamenteel Onderzoek der Materie (FOM)", which is financially supported by the "Nederlandse organisatie voor Wetenschappelijk Onderzoek (NWO)" and is part of an industrial partnership program between Philips and FOM.

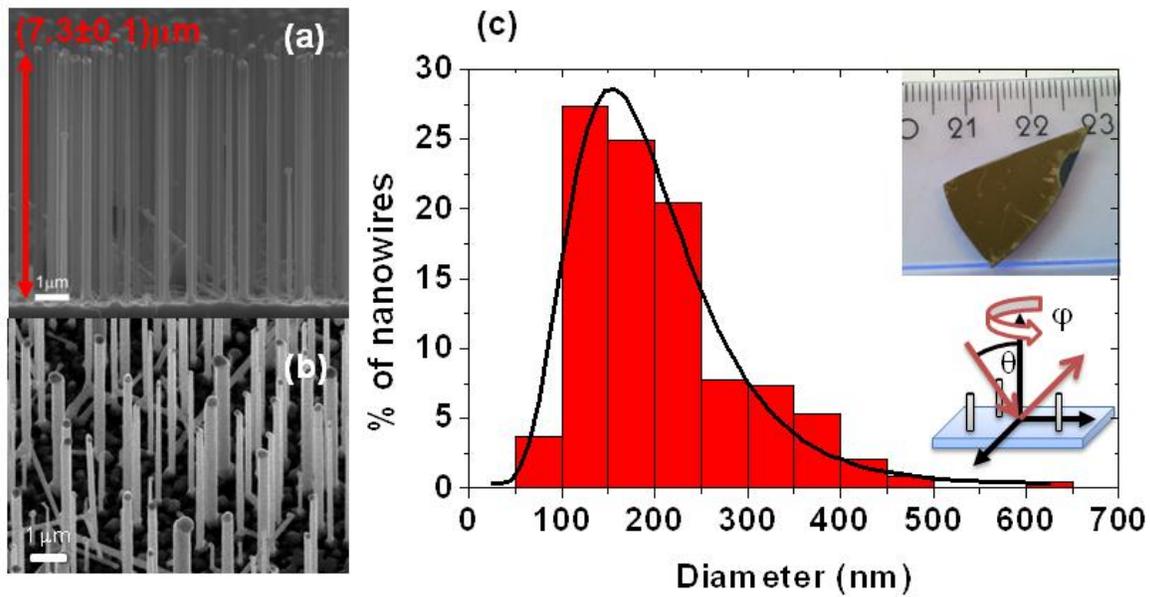

**FIG. 1.** Silicon nanowires epitaxially grown on a silicon substrate. (a) Scanning electron microscope (SEM) image of the cross section of the sample. (b) SEM image of the sample at an inclination of 30°. (c) Histogram of the percentage distribution of nanowire diameters approximated with a log-normal probability distribution (black solid line). The upper inset in (c) is a photograph of the sample placed on a ruler. The lower inset displays a schematic representation of the specular reflectance measurement where the angles θ and φ are the angle of incidence and the azimuthal angle of the incident beam onto the sample, respectively.

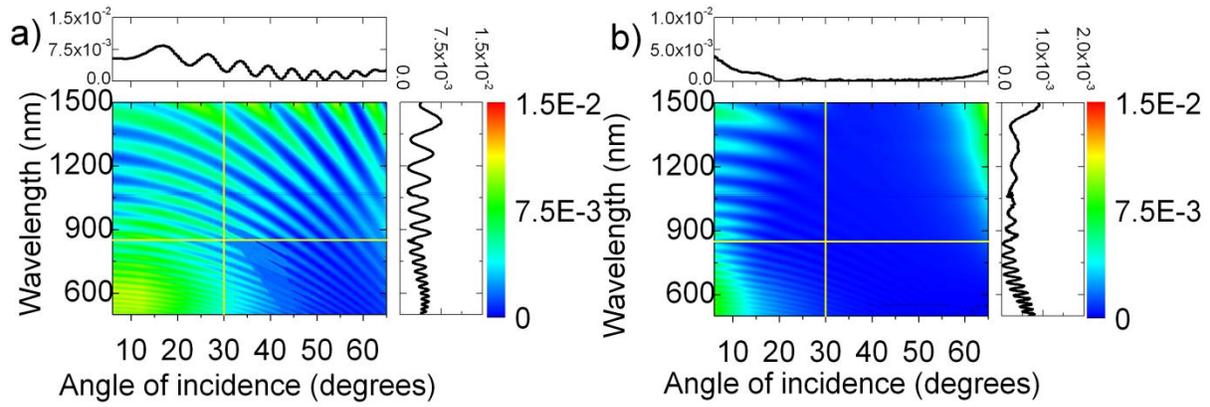

**FIG. 2.** Measured specular reflectance as a function of the wavelength and angle of incidence for (a) s-polarized and (b) p-polarized light. The graphs on the top and right sides of each figure are the specular reflectance at $\lambda$=850 nm and 30°, respectively.

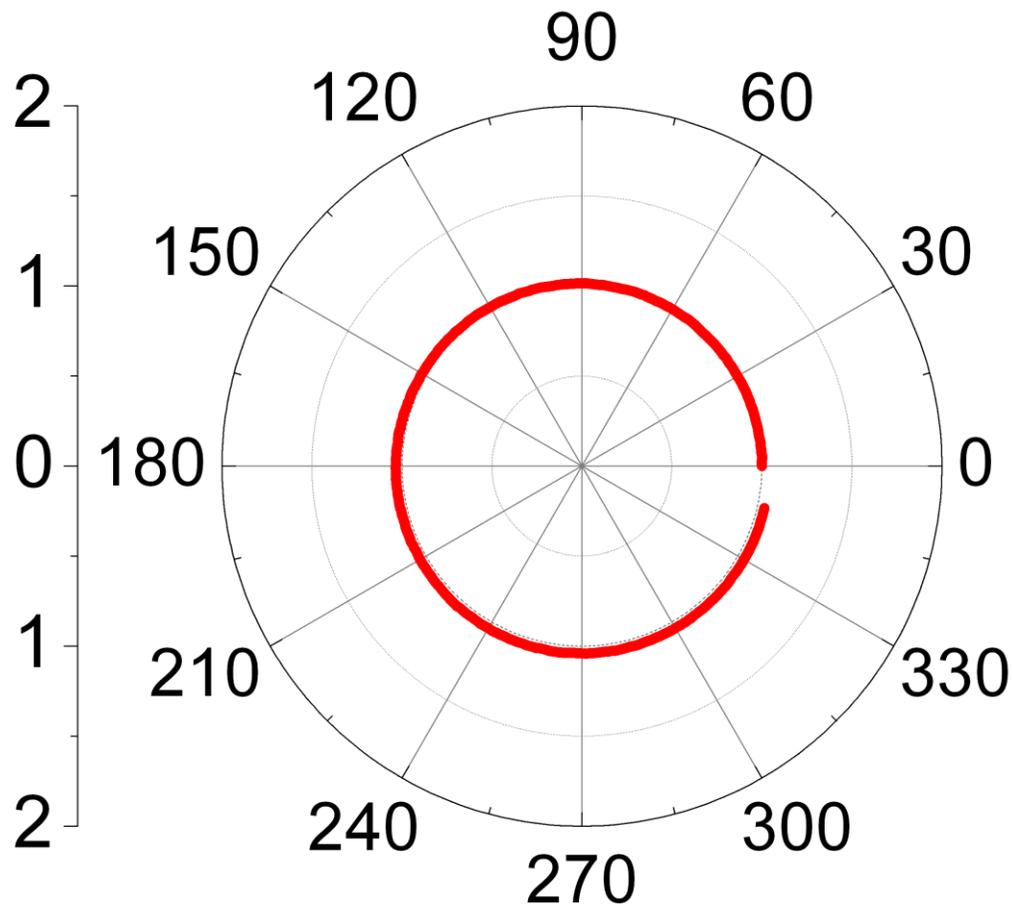

**FIG. 3.** Polar plot of the normalized specular reflectance as a function of the azimuthal angle measured at normal incidence with a HeNe laser ($\lambda$=633 nm). The sample was rotated along the axis parallel to the incident beam with steps of 1°. The reflectance is normalized to the value at $\varphi$ = 0°.

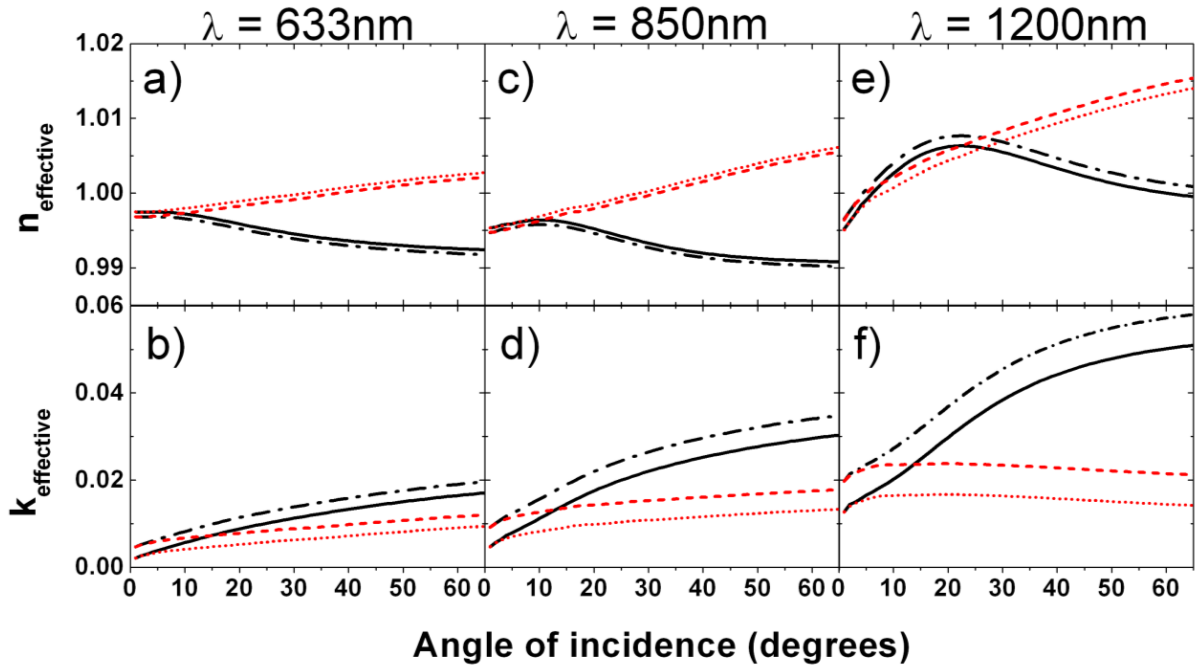

**FIG. 4.** Calculated effective refractive indices (a, c, and e) and extinction coefficients (b, d, and f) for the coherent beam in a polydisperse ensemble of silicon nanowires for three wavelengths: λ=633 nm (a) and (b), λ=850 nm (c) and (d), and λ=1200 nm (e) and (f). The calculations for p-polarized (black solid lines) and s-polarized (red dotted lines) light consider only vertical nanowires with a volume filling fraction $\alpha_{vertical} \cong 0.02$, whereas the calculations for vertical nanowires with $\alpha_{vertical} \cong 0.02$ and nonvertical nanowires with $\alpha_{nonvertical} \cong 0.005$ for p- and s-polarized light are represented by the black dash-dotted lines and the red dotted lines, respectively.

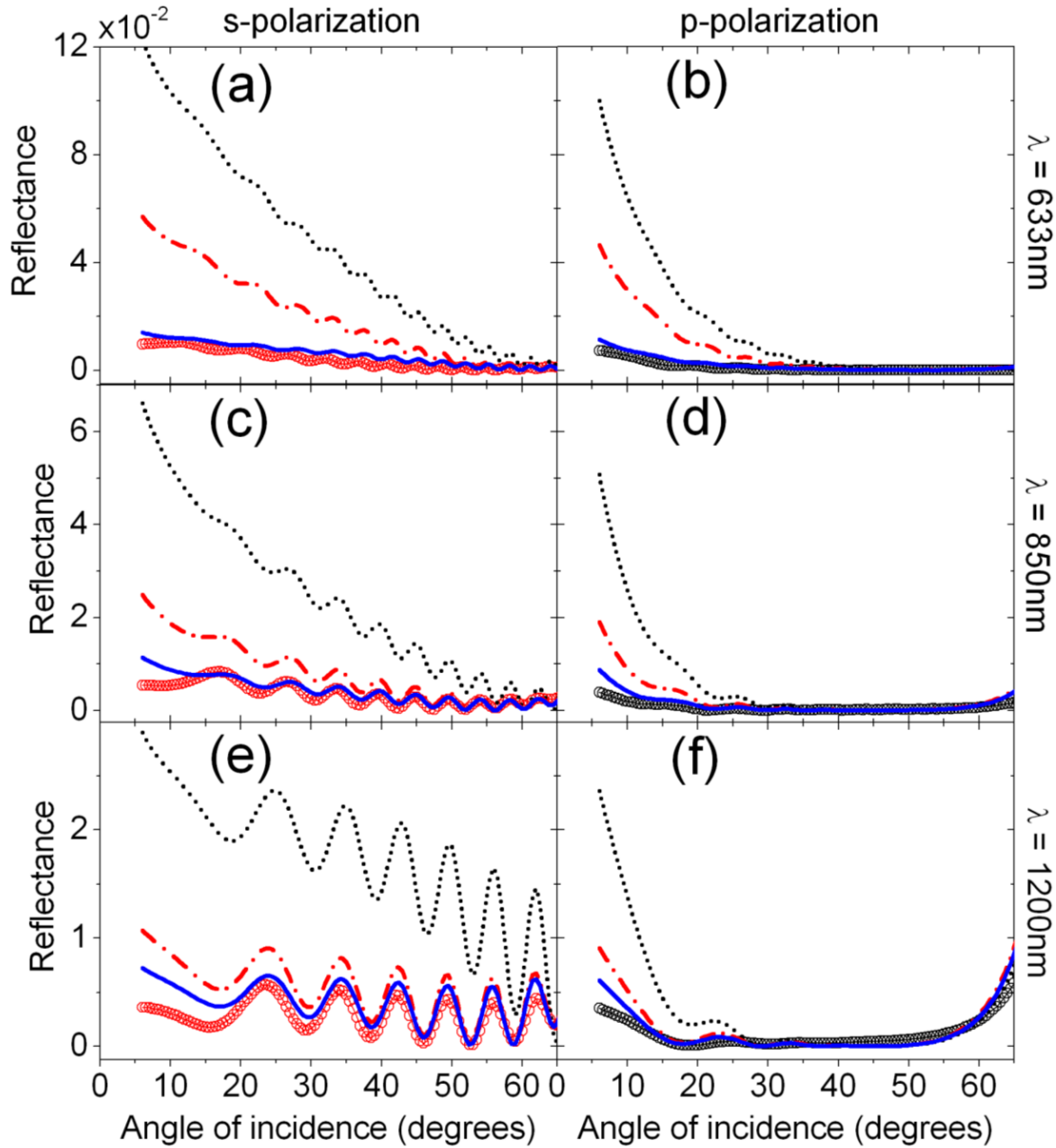

**FIG. 5.** Measured and calculated specular reflectance of s- and p-polarized light as a function of the angle of incidence for three different wavelengths (a and b) $\lambda$=633 nm, (c and d) $\lambda$=850 nm, and (e and f) $\lambda$=1200 nm). The measurements are displayed with circles (red for s-polarization and black for p-polarization). The black dotted lines represent the reflectance calculated for vertical nanowires, the red dash-dotted lines show the calculation including nonvertical nanowires and the blue solid lines represent the reflectance corrected for the surface scattering according to Eq. (8), with a root mean square of the roughness feature height of σ=60 nm.

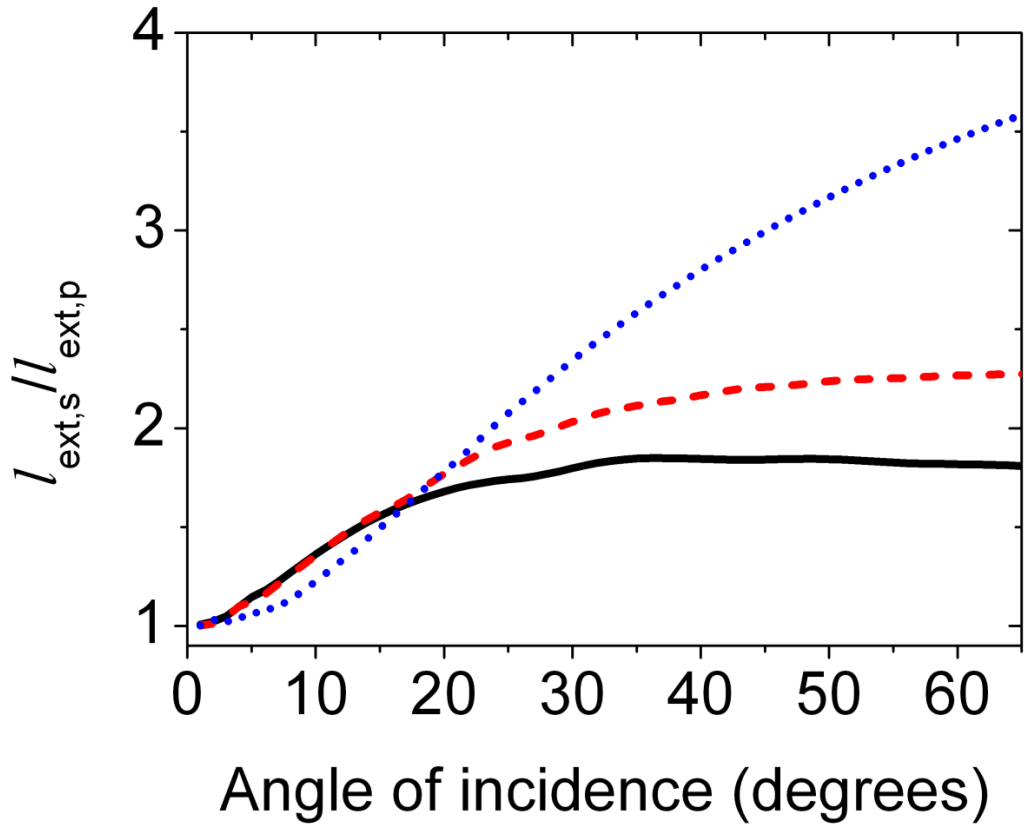

**FIG. 6.** Extinction anisotropy of a polydisperse ensemble of vertical nanowires represented as the ratio of the extinction lengths for s- and p-polarized light as a function of the angle of incidence. The extinction anisotropy is plotted for λ=633 nm (black solid line), λ=850 nm (red dashed line) and λ=1200 nm (blue dotted line).